# Electroluminescence of NV Color Centers in Diamond p-i-n Diodes mediated by Charge-state Dynamics

Ruirong Bai,[1,2] Menglin Huang,[2] Shanshan Wang,[2] Shiyou Chen[2,*] and Yu-Ning Wu[1,†]

[1] Key Lab of Polar Materials and Devices (MOE) and Department of Electronics, East China Normal University, Shanghai 200062, China

[2] College of Integrated Circuits & Micro-Nano Electronics and Key Laboratory of Computational Physical Sciences (MOE), Fudan University, Shanghai 200433, China

As the electroluminescence (EL) of NV color centers in diamond has been realized in p-i-n diodes, the underlying mechanism remains a puzzle for longer than a decade. In this study, using first-principles approaches, the electronic configurations and the possible transitions are comprehensively investigated. Based on the calculated carrier cross sections and transition rates, the mechanism of the EL of NV centers and the charge-state dynamics are revealed. The continuous EL is maintained by the cycle of $NV^0$ ground ($G_{NV^0}$), $NV^+$ metastable ($M_{NV^+}$) and $NV^0$ excited state ($E_{NV^0}$). The weaker EL intensity compared to photoluminescence (PL) is explained by the bottleneck transition from $M_{NV^+}$ to $E_{NV^0}$ and another non-luminescent transition cycle. Additionally, our results also explain the disappearance of the luminescence of $NV^-$ as a result of unbalanced transitions between $NV^-$ and $NV^0$. This study not only reveal the mechanism of electroluminescence of NV centers and explain experimental observations, but also provide first-principles insights to understand the charge-dynamics of other color centers under electric and optical field.

Color centers implemented in diamond- and silicon carbide-based p-i-n diodes can act as single-photon sources (SPSs) under electrical pumping[1-5]. They not only possess advantages such as higher integration, scalability and energy efficiency over the conventional optically pumped SPSs[6-8], but also become important platforms for charge-state control of the color center under electric and optical fields[3,9-14]. The electroluminescence (EL) of nitrogen vacancy (NV) centers was first realized in 2011[15]. Interestingly, only $NV^0$ luminescence is observed in such diodes, whereas both $NV^0$ and $NV^-$ signals are observed under optical pumping. Furthermore, interesting functions have been realized based on the electrical pumping of NV centers, including the deterministic initialization of the $NV^0$ charge state and the long afterglow after a sequential combination of optical and electrical pumping[3]. However, the photon emission rate of EL of NV centers is 1-2 orders lower than that of optical pumping[2,3,16,17], and also lower than other optically pumped color centers, including SiV[18] in diamond and $C_{Si}V_C$ in 4H-SiC [19].

To understand the experiments, Fedyanin et al[9,20-22] proposed a three-state cyclic model that the ground state and excited state of $NV^0$ are bridged by the $NV^-$ ground state. Specifically, the $NV^0$ ground state capture an electron from conduction band (CB) and transits to the $NV^-$ ground state, which further captures a hole from valence band (VB) and turns to $NV^0$ excited state. It emits a photon and deexcites to $NV^0$ ground state, forming a three-state cycle. This phenomenological model, however, assumes that the $NV^-$ ground state helps to form the cycle without considering other charge states[23], and the carrier capture cross sections (CCSs) used in this model are empirical. Mizuochi et al proposed another three-level model, in which the ground state of $NV^0$ is directly electrically pumped to a high-energy excited state (S = 3/2), followed by spin-orbit coupled transition (or intersystem crossing, ISC) to the $NV^0$ excited state and the radiative emission. Notably, this model omits the carrier capture process and the other charge states[2]. In fact, the continuous excitation and luminescence of NV centers in EL needs the carrier capture process to fulfill, which is non-trivial in phenomenological models. As a result, the mechanisms of the EL of NV centers remain not fully understood and uncertain, and experiments such as the weaker luminescence are not explained.

To reveal the mechanism of the EL of NV centers, we investigate the charge-state dynamics based on the first-principles evaluations of radiative/nonradiative transitions involved in the EL. Different from the reported models, multiple charge states are comprehensively considered, and the CCS and transition lifetime are calculated using first-principles approaches. Via comparing the CCS and lifetime of possible transitions, $NV^+$ metastable state, instead of $NV^-$ ground state, is found to be the bridge between the excited and ground states of $NV^0$. Furthermore, our model of transition paths also rationalizes multiple experiments unexplained by other models, including the weaker EL rate than the rate under optical pumping and the disappearance of the $NV^-$ in EL. The results are important for understanding and enhancing the EL of NV centers in diamond, and the approach paves a new path to disclosure the charge-state dynamics of color centers under electric and optical field.

The density functional theory (DFT)[24] as implemented in Vienna Ab-initio Simulation Package (VASP)[25] was used to perform the first-principles investigations. The HSE06 hybrid functional[26,27] and spin polarization were employed for defect and capture calculations. The constrained-DFT (CDFT) method[28] was employed to simulate the excited and metastable states. The formation energy and transition level of defect were calculated[29,30] together with the FNV correction method[30,31]. The nonradiative[32,33] and ISC transition rates[34] are calculated using quasi one-dimensional method, and the radiative[35] transition rate is obtained using the approach proposed by Dreyer et al[35]. The carrier capture cross section (CCS) can be calculated based on the carrier thermal velocity and carrier capture coefficient. The transition lifetimes are treated differently for the transition between two defect levels and the one between a defect level and the band edge, which are independent and dependent on the carrier concentration, respectively. More details can be found in Supplemental Material (SM), S1[36].

*Contact author: chensy@fudan.edu.cn

†Contact author: ynwu@phy.ecnu.edu.cn

The band gap of diamond is calculated as 5.40 eV, which is consistent with the experimental (5.48 eV[37]) and theoretical values (5.36-5.42 eV[37,38]). The formation energies and transition levels of four charge states, including $NV^{2-}$, $NV^{-}$, $NV^{0}$ and $NV^{+}$, are investigated, and the formation energies as functions of the Fermi level are shown in Figure 1(a). The transition levels of (0/+), (-1/0) and (-2/-1) are 0.98, 2.66 and 4.95 eV above valence band maximum, respectively, agreeing with the reported results (0.94, 2.70 and 4.90 eV) [23]. All these charge states may participate the EL, while the reported models at most consider $NV^0$ and $NV^-$ [20]. Additionally, besides the ground states, the excited states and metastable states should also be considered[2,22,39]. In fact, the excited states have been found to be crucial for nonradiative processes in wide-bandgap semiconductors[40].

Figure 1(b)-(d) present the considered electronic configurations for the ground states, metastable states and excited states for all charge states. The NV centers have six defect levels in the band gap with three spin-up ($a_1$, $e_x$ and $e_y$) and three spin-down ($\bar{a}_1$, $\bar{e}_x$ and $\bar{e}_y$) levels, respectively. As suggested by previous studies, some states may be multi-determinant states[41-46], and they are approximated by mixture occupancies [47] based on configuration interaction calculations[42](SM, S2).

For $NV^0$, the ground state ($G_{NV^0}$) has $^2E$ symmetry and its electronic configuration selected as $a_1\bar{a}_1e_x$ (left panel of Fig. 1(b))[13,42,48,49]. The excited states ($E_{NV^0}$) has $^2A_2$ symmetry and the electron occupation is mixed as $0.32\,a_1\,0.84\,(e_xe_y)\,0.66\,\bar{a}_1\,0.17\,(\bar{e}_x\bar{e}_y)$ (right panel of Fig. 1(b)). The considered metastable state ($M_{NV^0}$) with $^4A_2$ symmetry exhibits spin-flip with three spin-up orbitals $a_1e_xe_y$ occupied (middle panel of Fig. 1(b)).

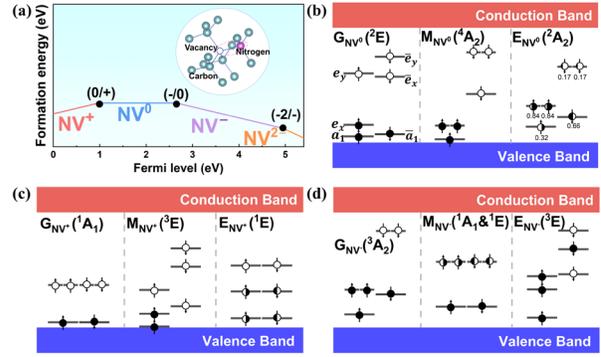

FIG 1. (a) The relative formation energies of the NV defect as functions of the Fermi level. (b)-(d) The considered electronic configurations of the ground state, metastable state and excited state for neutral, +1 and -1 charge states, respectively. The arrows up (down) indicate spin-up (spin-down) electronic orbitals, and solid (hollow) circles means the level is occupied (unoccupied). The numbers indicate how many electrons are occupied.

Similarly, the ground state of $NV^+$ ($G_{NV^+}$) exhibits $^1A_1$ symmetry and has $a_1\bar{a}_1$ occupation (left panel of Fig. 1(c)), and the metastable state ($M_{NV^+}$) with the $^3E$ symmetry and flipped spin has $a_1e_x$ occupation (middle panel of Fig. 1(c)). Two degenerate $^1E$ excited states ($E_{NV^+}$) exit[41,50] and the 0.5 ($a_1\bar{a}_1e_x\bar{e}_x$) mixed occupation was chosen (right panel of Fig. 1(c)).

As discussed in previous studies[42,43,45,46,49], the ground state of $NV^-$ ($G_{NV^-}$) has $^3A_2$ symmetry, while the excited state ($E_{NV^-}$) has $^3E$ symmetry and the metastable states ($M_{NV^-}$) with $^1E$ and $^1A_1$. For the $G_{NV^-}$ and $E_{NV^-}$, the occupations are $a_1\bar{a}_1e_xe_y$ (left panel of Fig. 1(d)) and $a_1e_x\bar{e}_xe_y$ (right panel of Fig. 1(d)), respectively[41,46,51]. The metastable states $^1E$ and $^1A_1$ are also multi-determinant states[41,46,51]. The $a_1\bar{a}_1$ $0.5(e_x\bar{e}_xe_y\bar{e}_y)$ occupation are chosen to approximate the $^1A_1$ and $^1E$ metastable states (middle panel of Fig. 1(d)). $NV^{2-}$ is also considered in this study (SM, S2[36]). However, it is found to be irrelevant in the EL (SM, S4[36]).

Based on the occupations in Fig. 1, the possible carrier capture processes, which can happen either non-radiatively or radiatively, are determined. Only the transitions occurring through capturing one carrier are considered, and transitions requiring multiple carrier capture processes are omitted, e.g. $M_{NV^0}$ to $G_{NV^+}$ transition needs two holes and one electron captures (SM, Fig. S2(b)[36]). Transitions involving multi-determinant states need to ignore higher order terms involving multi-carrier transitions (SM, S4[36]). Inspired by Alkauskas et al.[40], the possible transition paths can be integrated to the diagram of formation energies, which includes the metastable and excited states (Figure 2). All possible transitions are illustrated by the intersections of two formation energy lines with one charge difference (black dots), whereas the unmarked intersections means that the transitions cannot occur through single carrier capture.

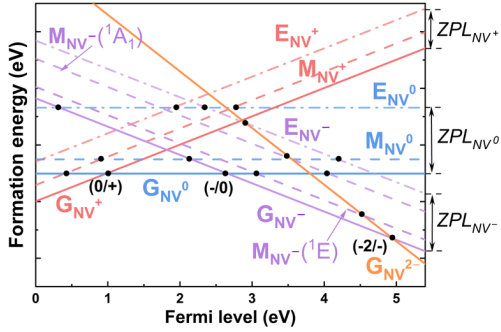

FIG 2. Formation energies of the NV centers with different charge states and electronic occupations as functions of the Fermi energy. The excited and metastable states are integrated so that the possible transitions via single carrier capture can be illustrated by the intersections. Unmarked intersections cannot occur via single-carrier capture (details in SM, S4[36]). The ZPLs are also labelled on the right-hand side.

Besides the carrier capture, there also exist transitions without changing the charge state, including the radiative and ISC transitions. The radiative transitions happen between the excited states and ground states. The zero-phono lines (ZPLs) of $NV^-$, $NV^0$ and $NV^+$ are calculated as 2.03, 2.32 and 1.40 eV, respectively (Fig. 2 and Table S1 of SM[36]). The ZPLs of $NV^-$ and $NV^0$ are in good agreement with the experiment (1.945 eV and 2.156 eV, respectively [49]). The calculated ZPL of $NV^+$ is matches the 1.60 eV calculated using the time-dependent density functional theory[50]. Based on the calculations of the lifetime of these three radiative transitions, they are all predicted to be bright states (SM, S3 and Table S3[36]). Indeed, $NV^-$ and $NV^0$ are observed to be bright states in experiments[2]. However, the luminescence of $NV^+$ have not been reported yet.

After discussing the different possible transitions, we calculate the cross sections for the radiative and nonradiative capture processes (Table S7[36]). By sorting the transitions with relatively higher CCSs, the charge-state dynamics during the EL of NV centers can hence be revealed. Starting with the transitions between $NV^-$ and $NV^0$, the non-radiative transitions mostly exhibit smaller CCSs than radiative processes, meaning that radiative transitions from band edges to defect levels have higher possibility to occur. This is because the wide bandgap of diamond generally elevates the potential barriers for non-radiative transitions between different charge states (SM, Fig.S1(b)[36]). The transitions from $NV^-$ to $NV^0$ has higher rate than the reverse processes. As shown in Table S4[36], the radiative CCSs (R-CCSs) from $G_{NV^-}$ to $G_{NV^0}/M_{NV^0}$ are all in the order of $10^{-20}$ cm$^2$, while the R-CCSs for the reverse processes are in the order of $10^{-21}$-$10^{-22}$ cm$^2$, which is 1-2 orders lower. The nonradiative CCSs (NR-CCSs) from $G_{NV^-}$ to $E_{NV^0}$ is in the order of $10^{-13}$ cm$^2$, which is seven orders larger than the R-CCS and the R-CCS of reverse process. Similar situations also apply for the transitions from $M_{NV^-}$ to $G_{NV^0}$ as well as from $E_{NV^-}$ to $M_{NV^0}/E_{NV^0}$. Such unbalanced transitions between $NV^-$ and $NV^0$ eventually lead to the depletion of $NV^-$, in particular the nonradiative process from $G_{NV^-}$ to $E_{NV^0}$, which explains the experiments that the no EL of $NV^-$ is observed[2]. Similar unbalanced transitions also exist in the transition between $NV^{2-}$ and $NV^-$, in which the $NV^{2-}$ transits faster to $NV^-$ than the revers process (Table S7 of SM[36]).

Further analysis of the transitions between $NV^0$ and $NV^+$ can reveal the mechanism of the continuous luminescence of $NV^0$. Interestingly, transitions from $NV^0$ to $NV^+$ states mostly prefer nonradiative multi-phonon carrier capture processes to radiative ones. Starting with $G_{NV^0}$, it has significantly high NR-CCSs (2.09×10$^{-15}$ cm$^2$) to transit to $M_{NV^+}$ via capturing holes from the VB (Fig. 3(a)), while the NR-CCS of $G_{NV^0}$ transiting to $G_{NV^+}$ is much lower (8.60×10$^{-21}$ cm$^2$). As a result, $G_{NV^0}$ mostly transits to $M_{NV^+}$ at rather fast rates. On the contrary, the reverse processes prefer radiative electron capture, and their R-CCSs are all in the order of $10^{-19}$-$10^{-20}$ cm$^2$, meaning that the $NV^+$ states turn to $G_{NV^0}$ states in slower rates.

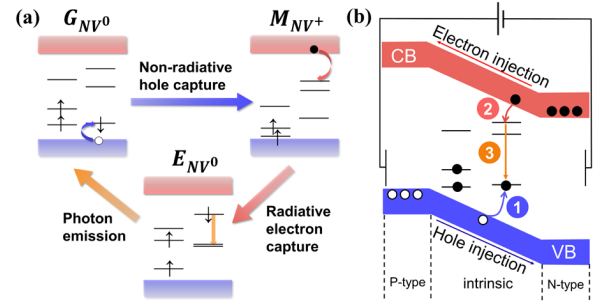

FIG 3. (a) Transition paths involved in EL of $NV^0$. $G_{NV^0}$ captures a hole nonradiatively and turn to $M_{NV^+}$. $M_{NV^+}$ captures an electron radiatively and transits to $E_{NV^0}$. $E_{NV^0}$ transits to $G_{NV^0}$ by emitting a photon. The occupation of $E_{NV^0}$ here is a case in the multi-determinant state. (b) Schematic illustration of EL of $NV^0$ center in a p-i-n diode via hole capture, electron capture and radiative process.

In terms of $M_{NV^+}$, there exist two possible transition paths. On one hand, it can transit to $E_{NV^0}$ through radiatively capturing one electron with R-CCSs 8.03×10$^{-19}$ cm$^2$ (Fig. 3(a)), and $E_{NV^0}$ will further illuminates and transits back to $G_{NV^0}$ (Fig. 3(a)). On the other hand, $M_{NV^+}$ can transit to $G_{NV^+}$ through ISC process. This path does not contribute to the EL of $NV^0$, because $G_{NV^+}$ cannot directly transit to $E_{NV^0}$.

The mechanism of EL of $NV^0$ are hence clear as demonstrated in Fig. 3(b). The p-type region and n-type region provide holes and electrons for the intrinsic region, respectively, where the NV$^0$ centers are located. The NV$^0$ ground state first changes to $M_{NV^+}$ by capturing a hole from VB, and subsequently transit to $E_{NV^0}$ by capturing an electron from CB. Eventually, $E_{NV^0}$ emits a photon and transits back to $G_{NV^0}$, forming a complete the cycle. This is fundamentally different from the reported three-state cycling model, in which $G_{NV^-}$ acts as the bridge between $G_{NV^0}$ and $E_{NV^0}$ [9,21,22]. We also examined the CCSs between different states involving EL cycle, validating the choice of configurations (SM, S4[36]).

We subsequently calculate the lifetime of the main transitions based on the capture coefficient at 300 K as shown in Fig. 4(a). The average nonequilibrium carrier concentration in the drift (intrinsic) region is set as 4.56×10$^{17}$ cm$^{-3}$ according to experiments and other parameters can be found in SM, S5[36]. The $G_{NV^0}$ transits to $M_{NV^+}$ rather fast through non-radiative process with lifetime 0.10 ns, followed by the radiative transition from $M_{NV^+}$ to $E_{NV^0}$ with lifetime 0.22 μs. Afterwards, the EL occurs as $E_{NV^0}$ radiatively transits to $G_{NV^0}$ with lifetime 18.4 ns. This cycle maintains the continuous luminescence of the NV$^0$ color center. However, it is bottlenecked by the slowest $M_{NV^+}$ to $E_{NV^0}$ transition, resulting in the weaker EL than PL.

$M_{NV^+}$ can also transit to $G_{NV^+}$ through ISC process with a lifetime of 0.04 ns. $G_{NV^+}$ can subsequently transit back to $G_{NV^0}$ by radiatively capturing electrons from CB with lifetimes of 1.40 μs. This cycle is bottlenecked by the slowest $G_{NV^+}$ transit to $G_{NV^0}$. The existence of the cycle also contributes to the weak EL intensity. Additionally, the afterglow exceeding 0.45 seconds after electrical pumping[3] can be understood as the result of the accumulation of $G_{NV^+}$ state caused by the bottleneck transitions.

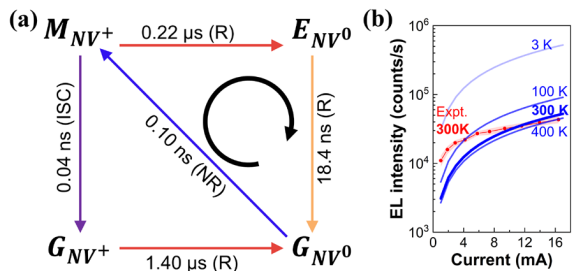

FIG 4. **(a)** Schematic of the main transition paths in the p-i-n diamond-based diodes with NV center. The blue arrow means the nonradiative hole capture process, and orange and red arrows are radiative transitions with or without changing the charge state, respectively. The ISC transition is indicated by purple arrow. The lifetimes are labeled on the arrow, and the abbreviations in the parentheses mark the transition types. The EL of NV$^0$ is maintained by the cycle indicated by the black arrow. **(b)** Comparison of the predicted EL rate (blue) of $NV^0$ with the experimental photon counts per second (red) for an individual NV[2].

We further calculate EL intensity (counts/s) of NV$^0$ centers as functions of the current at different temperatures based on the model shown in Fig. 4(a). Details of the EL intensity calculations can be found in S5 of SM. As shown in Fig. 4(b), at 300K, for currents ranging from 4 to 16 mA, emission rates are in the same order as the experimental data[2]. The rate is underestimated at lower currents probably because carrier from other shallow-level defects may enhance the carrier concentration and the formation of $E_{NV^0}$ [52,53], whereas the overestimation at higher currents may be an temperature-induced saturation of EL intensity in p-i-n diodes[2].

Furthermore, the emission rate is predicted to decrease as the temperature increases, different from the prediction by Fedyanin et al[22]. Although the increasing temperature may enhance the nonradiative carrier capture processes, the bottleneck transition of the EL cycle from $M_{NV^+}$ to $E_{NV^0}$, which is a radiative process, is less sensitive to temperature. In fact, the decreasing rate at higher temperature is mainly caused by the lower Sommerfeld factor (SM, Eq. (7) [36]), which means that the Coulombic interaction between the carriers and the defect charge state is suppressed and the recombination of free excitons is accelerated[54], thereby reducing the nonequilibrium carrier concentration and decreasing the EL rate. Therefore, lowering the working temperature can boost the bottleneck transition and enhance the EL intensity. For the same reason, the NV center close to the n-region is expected to exhibit higher EL intensity due to the higher electron concentration. Additionally, EL intensity is also expected to be enhanced by adding an external optical field that re-excites the $G_{NV^+}$ to $E_{NV^+}$, which reinforce the concentration of $E_{NV^+}$ and promote the transition of $E_{NV^+}$ to $E_{NV^0}$.

Last but not least, our results not only reveal the mechanism of EL, but also provide insight to understand the photoluminescence (PL) experiments. For example, in optical field modulation of the charge state experiments, only NV$^0$ luminescence is detected when irradiated with 637 nm laser[12]. Interestingly, the energy of 637 nm laser corresponds to the ZPL of NV$^-$ (1.945 eV), which is not high enough to excite $G_{NV^-}$ to $E_{NV^-}$, so no NV$^-$ luminescence is observed. However, the ZPL of $NV^0$ is 2.156 eV, even higher than that of NV$^-$, leaving the mechanism of NV$^0$ luminescence a mystery. Note that the calculated excitation energy of NV$^+$[42,50] is significantly lower than the energy of the 637 nm laser, so the $G_{NV^+}$ can be excited to $E_{NV^+}$. Meanwhile, the laser also induces carriers in the band edge by the shallow defect excitations or two-photon process excitations of NV centers. Based on our model, $E_{NV^+}$ has two paths to reach $E_{NV^0}$, and then to $G_{NV^0}$ by photon emission. $E_{NV^+}$ can radiatively transits to $E_{NV^0}$ through electron capture (SM, Tab. S7[36]), or it can reach $M_{NV^+}$ through ISC process, and then radiatively capture an electron to $E_{NV^0}$. $G_{NV^0}$ nonradiatively captures a hole and transit to $M_{NV^+}$, which then initiates the luminescence cycle in Figure 4(a). Importantly, different from EL, 637 nm laser can efficiently excite the $G_{NV^+}$ back to $E_{NV^+}$. It pushes the $E_{NV^+}$ to transit to $E_{NV^0}$, maintaining the luminescence cycle.

To conclude, this study reveals the EL mechanism of NV centers and charge-state dynamics, and explains multiple experiments. The continuous EL is maintained by a three-state cycle among $G_{NV^0}$, $M_{NV^+}$ and $E_{NV^0}$. The weaker EL intensity compared to PL is caused by the bottleneck transition from $M_{NV^+}$ to $E_{NV^0}$ and another transition cycle among $G_{NV^0}$, $M_{NV^+}$ and $G_{NV^+}$. The absence of NV⁻ EL is due to the depletion of NV⁻ caused by the unbalanced transition rate between $NV^0$ and NV⁻ states. The long afterglow of $NV^0$ comes from the accumulation of $G_{NV^+}$ states caused by the slow transition from $M_{NV^+}$ to $E_{NV^0}$. Our study clarifies the diamond NV center EL mechanism and supports efforts to improve the efficiency of NV center single-photon sources. This study reveals the mechanism of the EL of NV centers and explains the experiments puzzled for a decade. More importantly, our approach paves a new path to disclose the charge-state dynamics of color centers under electric and optical field.

This work was supported by the National Key Research and Development Program of China (2022YFA1404603 and 2022YFA1402904), the National Natural Science Foundation of China (NSFC) under grant Nos. 12174060, 12204170, 12474069 and 12334005, the Science and Technology Commission of Shanghai Municipality (24JD1400600), the Shanghai Municipal Commission of Education (the AI Empowered Research Program 2024AI02001) and Project of MOE Innovation Platform.